# Influence of strain on the anomalous Hall and Nernst effects in Fe thin films

Ao Nakagawa[1,2,3], Ryo Toyama[2], Keisuke Masuda[2], Weinan Zhou[2], Hirofumi Suto[2], Kodchakorn Simalaotao[1,2], Yoshio Miura[2,4], Yuya Sakuraba[1,2], Tetsunori Koda[3]

[1]Graduate School of Science and Technology, University of Tsukuba, Tsukuba, Ibaraki, Japan

[2]Research Center for Magnetic and Spintronic Materials (CMSM), National Institute for Materials Science (NIMS), Tsukuba, Ibaraki, Japan

[3]Electronic Mechanical Engineering Department, National Institute of Technology, Oshima college, Suo-Oshima, Yamaguchi, Japan

[4]Faculty of Electrical Engineering and Electronics, Kyoto Institute of Technology, Sakyo-ku, Kyoto, Japan

## Abstract

  The anomalous Hall effect (AHE) and anomalous Nernst effect (ANE) are the transverse transport phenomena in magnetic materials, which reflect the Berry curvature arising from the electronic structure near the Fermi level. Lattice strain provides a direct means to tune these effects by modifying the electronic structure; however, disentangling the strain-induced effect through the Berry curvature modulations in multicomponent materials is challenging due to complexities arising from extrinsic contributions by impurities and disorder, as well as difficulties in simple direct comparison with first-principles calculations. In this study, we focus on Fe, a prototypical single element ferromagnet with a well-established electronic structure, and tune the sign and magnitude of the strain in epitaxial thin films of by varying the substrates and deposition conditions to investigate the strain effect on the AHE and ANE. Scaling law analysis revealed that the intrinsic anomalous Hall conductivity (AHC) exhibits a clear tetragonal distortion ($c/a$) dependence, in good agreement with theoretical calculations based on Berry curvature modification. In contrast, the anomalous Nernst conductivity (ANC) shows a pronounced deviation from the theoretical values and markedly different $c/a$ dependence. These results demonstrate a crucial difference in the physical origin between the AHC and the ANC in the Fe films; the AHC is predominantly governed by intrinsic mechanisms, whereas the ANC is strongly influenced by the extrinsic contribution.

Contact authors : SAKURABA.Yuya@nims.go.jp, koda@oshima-k.ac.jp

The electronic structure of a material plays a crucial role in determining its electrical, magnetic, and thermoelectric properties. Among these, the anomalous Hall effect (AHE) and the anomalous Nernst effect (ANE) are representative transverse transport phenomena in magnetic materials, both of which strongly depend on the electronic structure near the Fermi level ($E_F$). In many magnetic materials, the AHE and ANE originate from the Berry curvature arising from the electronic structure in the presence of spin-orbit interaction [1–10]. Recently, giant AHE and ANE driven by enhanced Berry curvature have been observed in $Co_2MnGa$ [7], $Co_2MnAl$ [9], $Fe_3Ga$ [8], and $Co_3Sn_2S_2$ [10] due to their topological band structures near the $E_F$. These findings provide not only fundamental insights into Berry curvature driven transport but also significant potential for innovative magnetic device applications [11–17].

Modulating the electronic structure near the $E_F$—through external fields, compositional tuning, or carrier doping—is expected to significantly alter the Berry curvature, thereby inducing substantial changes in the AHE and the ANE. Among these approaches, controlling the electronic structure via composition and carrier concentration has been the most extensively investigated [18–22]. For instance, in $Co_2MnGa$, pronounced variations in the AHE and the ANE have been observed in thin films with off-stoichiometric compositions achieved by tuning the composition ratios [18]. In contrast, systematic investigations on the modulation of the AHE and the ANE by external fields—particularly application of mechanical strain—remain limited. Given strong coupling between the electronic and crystal structures, together with the high sensitivity of the AHE and the ANE to Berry curvature near the $E_F$, strain is expected to induce remarkable changes in these effects. Indeed, epitaxial strain has been shown to tune the Berry curvature and modify the AHE in $SrRuO_3$ thin films grown on various single-crystal substrates [23], where the effect was carefully analyzed through the strain-induced various changes such as oxygen octahedral rotations and crystal-field splitting. While strain-induced variations in the AHE and the ANE have also been reported in Weyl antiferromagnet $Mn_3Sn$ films [24] and polycrystalline $Co_2MnGa$/AlN multilayer films [25], their origins were attributed to the strain-induced rotation of the antiferromagnetic order and enhancement of the Seebeck effect, respectively, rather than to a direct modification of the Berry curvature. Therefore, in multicomponent materials and those-based nanostructures, isolating the intrinsic strain-induced effects on Berry curvature is not always straightforward. This complexity arises because both the AHE and the ANE are affected by modulations in the Berry curvature induced not only by strain, but also by variations in atomic ordering, vacancy formation, off-stoichiometry, chemical bonding, and magnetic structure. The presence of extrinsic contributions such as side-jump and skew scattering due to impurities and defects further complicates the analysis.

Fe is the most fundamental ferromagnetic element and has been extensively employed to elucidate various phenomena in spintronics because its electronic structure can be analyzed rigorously free from the complexities in the multicomponent materials [26–32]. Therefore, investigating the strain-induced modifications of the AHE and the ANE in Fe provides a model platform to examine the modulation of Berry curvature near the $E_F$. More importantly, for the AHE, previous studies have established a scale law model between the longitudinal and anomalous Hall resistivities ($\rho_{xx}$ and $\rho_{yx}^A$), which enables us to extract only the intrinsic contribution from the experimentally obtained $\rho_{yx}^A$ containing certain extrinsic contributions [29,33–42]. Since these scaling analyses have been validated in pure Fe, their applicability has been already well justified [29,42]. As a control method for the strain, the epitaxial strain due to the mismatch between films and substrates is effective. High quality (001)-oriented epitaxial Fe films can be grown by choosing suitable single crystalline substrates [29,32,43,44], allowing controlled application of both in-plane tensile and compressive epitaxial strains via lattice mismatch.

Therefore, in this study, we systematically investigated the strain-induced effect on the AHE and the ANE in epitaxial Fe thin films grown on MgO or $MgAl_2O_4$ (MAO) substrates, which impose in-plane tensile and compressive strain, respectively. By analyzing the AHE using a scaling law, we find that the strain dependence of the intrinsic anomalous Hall conductivity (AHC) is in good agreement with our theoretical calculation, demonstrating that the strain-induced Berry curvature modulation on AHE in Fe. In contrast, the anomalous Nernst conductivity (ANC, $\alpha_{xy}^A$) exhibits a small negative value regardless of the magnitude and sign of strain, which is remarkably different from our theoretical prediction. This discrepancy indicates that extrinsic mechanisms play a significant role in the ANE in pure Fe, highlighting a fundamental difference in the origin of the AHE and the ANE under strain.

To systematically change the magnitude and direction of the strain, we prepared total six (100)-oriented Fe epitaxial films on single-crystal MgO(100) and $MgAl_2O_4$ (MAO)(100) substrates under different deposition conditions using two sputtering systems: ECS016 (EIKO) and CMS-A6250X2 (Comet, Inc.) (hereafter called EIKO-sputter and Comet-sputter, respectively.) as shown in Table I. The Fe films grown on the MgO (MAO) substrate is expected to have the lattice mismatch of +3.8% (-0.3%) using the lattice constants of bulk bcc-Fe (≈ 2.866 Å [45]), MgO (≈ 4.212 Å [45]), and MAO (≈ 8.086 Å [46]), giving an in-plane biaxial tensile (compressive) strain. The designed thickness of Fe thin films was fixed at 20 nm to preserve the strain induced by lattice mismatch from the substrate without a structural relaxation, while minimizing contributions from the surface and substrate interface to the transport properties. All films were deposited at room temperature using two different sputtering systems, each equipped with a different Fe target of the same purity (99.99%). Using EIKO-sputter, two Fe films were deposited by DC sputtering at an Ar gas pressure of 0.67 Pa. To modulate the strain via substrate surface conditions, two types of surface treatments were performed prior to the deposition: One Fe film underwent Ar-ion milling for 5 mins while the other underwent flash annealing at 650°C. Using Comet-sputter, four Fe films were deposited by RF sputtering at an Ar gas pressure of 0.29 Pa after flash annealing (600°C, 1 hour) on MgO and MAO substrates. To further modify the strain, half of the deposited Fe films (one on MgO and one on MAO) were subsequently post-annealed at 300°C for 30 min. To prevent the oxidization, all films were capped with 2nm-thick Al or Ta layer at room temperature. The crystal structure including a lattice distortion was characterized by the out-of-plane and tilted-plane x-ray diffraction (XRD) measurements with a Cu-Kα source at the tilted angle $\chi = 0°$ and $45.0°$. The actual thickness was measured by x-ray reflectivity (XRR). The magnetoresistance and the Hall resistivity ($\rho_{yx}$) were measured using Physical Property Measurement System (PPMS DynaCool; Quantum Design) with temperature range from 15 K to 300 K. The Seebeck electric field ($E_{SE}$) and the Nernst electric field ($E_y$) were simultaneously measured in PPMS Versalab (Quantum Design) at 300 K with varying the temperature gradient ($\nabla T$). $\nabla T$ was precisely measured by the patterned Au-wire on-chip thermometer made, as established in previous studies [47,48]. $H$ was applied to the out-of-plane direction of the substrate surface. Subsequently, the $\rho_{xx}$, $\rho_{yx}^A$, Seebeck coefficient ($S_{SE}$), and anomalous Nernst coefficient ($S_{ANE}$) were analyzed. The details of the measurement setups and analysis methods are mentioned in the Supplemental Material.

Table I: Deposition conditions for six prepared Fe epitaxial films. AD stands for "As-deposited".

| No. | Name | Substrate | Substrate cleaning | Ar gas pressure (Pa) | RF or DC | Sputtering system | Target-substrate distance (cm) | Post-annealing |
|---|---|---|---|---|---|---|---|---|
| 1 | AD-MgO-DC-R1 | MgO | Ar-ion milling (Room temperature, 5 min) | 0.67 | DC | EIKO | 20 | AD |

| | | | Flash annealing | 0.67 | DC | EIKO | 20 | AD |
| 2 | AD-MgO-DC-R2 | MgO | (650°C, 1 hour) | | | | | |
| 3 | AD-MgO-RF | MgO | Flash annealing (600°C, 1hour) | 0.29 | RF | Comet | 27 | AD |
| 4 | AD-MAO-RF | MgAl$_2$O$_4$ | Flash annealing (600°C, 1hour) | 0.29 | RF | Comet | 27 | AD |
| 5 | 300°C-MgO-RF | MgO | Flash annealing (600°C, 1hour) | 0.29 | RF | Comet | 27 | 300°C |
| 6 | 300°C-MAO-RF | MgAl$_2$O$_4$ | Flash annealing (600°C, 1hour) | 0.29 | RF | Comet | 27 | 300°C |

Figures 1(a) and 1(b) show the XRD profiles of Fe thin films at the tilt angle $\chi = 0°$ and 45.0°, respectively. In Fig. 1(a), the peak of 200 is detected at $2\theta \approx 65°$, indicating that (001)-oriented Fe thin films are grown on the substrates. The angle corresponding to 200 peak clearly shifted, suggesting that the lattice constants of the c-axis was affected by the substrates and the deposition conditions. In Fig. 1(b), the peak of 110 was observed at $2\theta \approx 45°$, indicating the (001)-oriented epitaxial growth. The c-axis and a-axis lattice constants and the *c/a* ratios, which are evaluated by 200 and 110 peak positions, are shown in Figs. 1(c) and 1(d). Without post-annealing, the Fe films on MgO substrates (AD-MgO-DC-R1, AD-MgO-DC-R2 and AD-MgO-RF) exhibit the in-plane tensile strain (*c/a* < 1), whereas the Fe film on the MAO substrate (AD-MAO-RF) exhibits the in-plane compressive strain (*c/a* > 1) as expected from the sign of lattice misfit. After the post-annealing, two Fe films (AD-MgO-RF and AD-MAO-RF) exhibit the nearly identical *c/a* ratio regardless of the substrates, (300°C-MgO-RF and 300°C-MAO-RF), which could be attributed to strain relaxation induced by the formation of the dislocation at the substrate interface. As a result, the *c/a* ratio of six Fe films spans in wide range from 0.9825 to 1.0108, with maintaining the (001)-oriented epitaxial structure. It is noteworthy, the shape of the 200 peak for AD-MgO-RF is asymmetrical with a tail towards the high-angle side. To quantitatively evaluate the asymmetry of the 200 peak shapes, we performed a fitting using the skewed Voigt model, which incorporates a skewness parameter ($\gamma$) into the standard Voigt distribution [49]. The inset of Fig. 1(d) shows the *c/a* ratio dependence of $\gamma$. The $\gamma$ value at *c/a* = 0.9778 (AD-MgO-RF) showed a finite value, while the $\gamma$ values of other Fe films were almost zero, indicating the presence of microscopic strain gradients only in AD-MgO-RF.

Therefore, to separately exhibit the result of this film from other films with uniform strains, the triangular symbol is used in all figures shown later.

The $H$ dependence of $\rho_{xx}$ and $\rho_{yx}$ were measured for all six Fe films from 15 to 300 K and obtained $\rho_{xx}$ and $\rho_{yx}^A$ were summarized with respect to $c/a$ ratios at each temperature (see Fig.S2 in Supplemental Material). To separate the intrinsic and extrinsic contributions to the AHE from these results, we employed a simplified scaling analysis [29]. The scaling equation is expressed as

$$\rho_{yx}^A = \alpha_{sk}\rho_{xx0} + b_{int}\rho_{xx}^2, \qquad (4)$$

where $\alpha_{sk}$ represents the extrinsic contribution of skew scattering, $b_{int}$ expresses the intrinsic AHC, $\rho_{xx0}$ is the residual resistivity ($\rho_{xx}$ at 15 K is used in this study), respectively [29]. The extrinsic contribution of side-jump is not considered because previous studies have confirmed it is negligibly small in comparison with the intrinsic contribution [50] and the subsequent studies demonstrate this assumption yields reasonable fitting results [34–36]. The $\rho_{xx}$ dependence of $\rho_{yx}^A$ and the fitting result with eq. (4) are shown in Fig. 2(a). No significant deviations are observed between the data points and the fitted lines, with coefficients of determination ($R^2$) greater than or equal to 0.9998 in all Fe films. The $c/a$ ratio dependence of $\alpha_{sk}$ is shown in Fig. 2(b). The sign of $\alpha_{sk}$ alters from negative to positive with the $c/a$ ratio, showing the variation in the contribution of the skew scattering among these Fe films. The origin of this $c/a$ ratio dependence of $\alpha_{sk}$ is still unclear. However, considering the source of skew scattering is the spin-dependent scatterings by the impurities [51], the variation of $\alpha_{sk}$ in these Fe films could be attributed to the amount of impurities rather than the $c/a$ ratio. In fact, a previous study has reported that $\alpha_{sk}$ is sensitive to even small amounts of impurities less than 0.5 at.% [35]. Since the two Fe films exhibiting the negative $\alpha_{sk}$ value were deposited in EIKO-sputter while the other films showing the positive $\alpha_{sk}$ value were deposited in Comet-sputter, we speculate that this sign reversal is caused by differences in the amount or type of impurities in the Fe films due to two Fe targets and environmental factors in these chamber. The $b_{int}$ value also clearly changes with the $c/a$ ratio as shown in Fig. 2(c). The $b_{int}$ value in the Fe film with almost cubic ($c/a$ ratio = 0.9983) is 1012 S/cm , which is in good agreement with the values for Fe thin films reported by Tian et al [29]. The origin of this $c/a$ ratio dependence will be discussed later based on the first-principles calculation of the intrinsic AHC ($\sigma_{xy}^{int}$). To evaluate the extrinsic contribution to the AHE, we reproduced $\rho_{yx}^A$ from $\alpha_{sk}$ and $b_{int}$ based on the scaling law and estimated the extrinsic contribution ratio at 300 K using $\alpha_{sk}\rho_{xx0} / \rho_{yx}^A$. The $c/a$ ratio dependence of the extrinsic contribution ratio is shown in Fig. 2(d). The magnitude of the extrinsic contribution ratio is less than 3% in all the Fe films, revealing that the extrinsic contribution to the AHE is almost negligible. It should be noted that the $\alpha_{sk}$ and $b_{int}$ values decrease rapidly in the sample with strain gradients, which suggests that microscopic strain gradients affect both the skew scattering and the electronic structure.

Figure 3(a) shows the $H$ dependence of $E_y$ normalized by $\nabla T$ for each *c/a* ratio. All curves exhibit the parabolic curve in the range from ≈ –2 to ≈ 2 T. Since the ANE in Fe is very small, the non-negligible even function component is overlapped with the signal of the ANE. This component arises from the planar Nernst effect (PNE), which originates from the anisotropic thermoelectric response depending on the relative angle between the in-plane component of magnetization and the applied temperature gradient. The *c/a* ratio dependence of $S_{SE}$ and $S_{ANE}$ are shown in Figs. 3(b) and 3(c). The $S_{SE}$ was positive ranging from ≈ 7.7 to ≈ 14.7 µV/K, while the $S_{ANE}$ was negative ranging from ≈ –0.15 to ≈ –0.30 µV/K. The negative $S_{ANE}$ values agree with the previous reports for the Fe films [17,52]. The $S_{ANE}$ can be decomposed into two terms as follows [53]:

$$S_{\mathrm{ANE}} = \rho_{xx}\alpha_{xy}^A - S_{SE}\left(\frac{\rho_{yx}^A}{\rho_{xx}}\right). \tag{5}$$

The first term ($S_1 = \rho_{xx}\alpha_{xy}^A$) and the second term ($S_2 = -S_{SE}\rho_{yx}^A/\rho_{xx}$) on the right-hand side of Eq. (5) represent the direct conversion of the thermal current into the transverse electric field and the conversion of a carrier flow driven by the SE into the transverse electric field via the AHE, respectively [53]. As shown in Fig. 3(c), both estimated $S_1$ and $S_2$ are negative, and the $S_2$ is more dominant for observed negative $S_{ANE}$ except for *c/a* ratio = 0.9825. Finally the $\alpha_{xy}^A$ (=$S_1/\rho_{xx}$) is evaluated and plotted with respect to the *c/a* ratio (Fig. 3(d)). The $\alpha_{xy}^A$ are negative in the samples with the uniform strain. Only sample with strain gradients exhibits a unique behavior; the $\alpha_{xy}^A$ is almost zero.

To analyze the experimental *c/a* ratio dependence of $b_{int}$ and $\alpha_{xy}^A$ in terms of intrinsic contribution, we performed first-principles calculations for the intrinsic AHC ($\sigma_{xy}^{int}$) and the intrinsic ANC ($\alpha_{xy}^{int}$) in Fe with and without strains. The lattice constants employed in the calculations are shown in Table II. These lattice constants were determined to vary the *c/a* ratios in increments of approximately 0.01 while keeping the volume approximately constant. The electronic structure was calculated based on the density functional theory (DFT) using the Vienna *ab initio* simulation program (VASP). From the calculation result, we estimated the energy dependence of $\sigma_{xy}^{int}$ expressed as

$$\sigma_{xy}^{int}(\varepsilon) = -\frac{e^2}{\hbar}\int\frac{d^3k}{(2\pi)^3}\,\Omega^z(\mathbf{k}), \tag{1}$$

$$\Omega^z(\mathbf{k}) = -\left(\frac{\hbar}{m}\right)^2 \sum_n f(E_{n,\mathbf{k}},\varepsilon) \sum_{n\neq n'} \frac{2\,\mathrm{Im}\langle\psi_{n,\mathbf{k}}|p_x|\psi_{n',\mathbf{k}}\rangle\langle\psi_{n',\mathbf{k}}|p_y|\psi_{n,\mathbf{k}}\rangle}{(E_{n',\mathbf{k}} - E_{n,\mathbf{k}})^2}, \tag{2}$$

where $e$ is the elementary charge of electron, $\hbar$ is the Planck constant divided by $2\pi$, $\Omega^z(\mathbf{k})$ is the Berry curvature, $m$ is the mass of electron, $n$ and $n'$ are the band indies, $p_x$ ($p_y$) is the $x$ ($y$) component of the momentum operator, $\psi_{n,\mathbf{k}}$ is the eigenstate with the eigenenergy $E_{n,\mathbf{k}}$, and $f(E_{n,\mathbf{k}},\varepsilon)$ is the Fermi distribution function for the band $n$ and the wave vector $\mathbf{k}$ at the energy $\varepsilon$ relative to the Fermi level. The 151×151×151 k-points were used, with the magnetization aligned along the [001] direction to match the experimental condition. Subsequently, the energy dependence of $\alpha_{xy}^{int}$ was calculated using

$$\alpha_{xy}^{int} = -\frac{1}{eT}\int d\varepsilon \left(-\frac{\partial f}{\partial \varepsilon}\right)(\varepsilon - \mu)\sigma_{xy}^{int}(\varepsilon), \tag{3}$$

where $T$ is the temperature ($T$ is fixed at 300 K in this study), $f = 1/\{\exp[(\varepsilon - \mu)/k_B T] + 1\}$ is the Fermi distribution function with $\mu$ being the chemical potential ($\mu = 0$ corresponds to the Fermi level).

Table II: Summary for lattice constants used in the theoretical calculation.

| Lattice constant $a$ (Å) | Lattice constant $c$ (Å) | $c/a$ ratio |
|---|---|---|
| 2.856 | 2.887 | 1.011 |
| 2.859 | 2.859 | 1.000 |
| 2.881 | 2.848 | 0.989 |
| 2.899 | 2.834 | 0.978 |

The $\mu$ dependence of $\sigma_{xy}^{int}$ and $\alpha_{xy}^{int}$ on each $c/a$ ratio are shown in Figs. 4(a) and 4(b). The energy dependence of $\sigma_{xy}^{int}$ and $\alpha_{xy}^{int}$ significantly varies with the $c/a$ ratio. In the case of $c/a = 1.000$ (normal cubic Fe), the peaks of $\sigma_{xy}^{int}$ appear around $\mu = -0.25, 0.0$, and 0.25 eV, which agrees with previous results by Weischenberg et al. [30]. The $\alpha_{xy}^{int}$ values at $\mu = 0$ eV range from –0.74 to 2.2 A/(m·K). Figures 4(c) and 4(d) show the $c/a$ ratio dependence of the experimental results ($b_{int}$ and $\alpha_{xy}^A$) and the theoretical results ($\sigma_{xy}^{int}$ and $\alpha_{xy}^{int}$) calculated at $\mu = 0$ eV. In the samples with uniform strains, the $b_{int}$ value is close to the theoretical $\sigma_{xy}^{int}$ and the $c/a$ ratio dependence of $b_{int}$ qualitatively agrees with that of $\sigma_{xy}^{int}$; both $b_{int}$ and $\sigma_{xy}^{int}$ reduce from the $c/a$ of around 0.980 to 0.987 and increases at around 0.998-1.000. This result indicates that the intrinsic contribution to the AHE in these Fe films alters by the strain-induced modulation of the Berry curvature near the $E_F$. It should be noted that the quantitative deviation between $b_{int}$ and $\sigma_{xy}^{int}$ can be attributed to the interfacial effect of electron scattering at the surface and the substrate in the Fe films, which cannot be eliminated by the analysis based on the scaling law.

By contrast, $\alpha_{xy}^A$ exhibits the negative values irrespective of the *c/a* ratio, in disagreement with the theoretical $\alpha_{xy}^{int}$ as shown in Fig. 4(d). We discuss the origin of this discrepancy as follows. For the AHE, extensive experiments and theoretical studies have proposed the scaling laws governing the intrinsic and extrinsic contributions linking $\rho_{xx}$ and $\rho_{yx}^A$, enabling to isolate the extrinsic component and evaluate only intrinsic AHC $b_{int}$. On the other hand, for the ANE, no established scaling law exists to separate intrinsic and extrinsic contributions using the longitudinal thermoelectric tensor ($\alpha_{xx}$). As a result, the measured $\alpha_{xy}^A$ in the present Fe films contains both intrinsic and extrinsic contributions. Indeed, the discrepancy between $\alpha_{xy}^A$ and $\alpha_{xy}^{int}$ has been reported in the previous studies, and several potential origins have been discussed, such as a shift in the $E_F$ due to off-stoichiometry or defects [18,54,55], atomic disorder [19], the effect of on-site Coulomb interactions [34,48], and extrinsic contributions [18]. However, in the present case, since Fe is a single-element material, the effects of off-stoichiometry and atomic disorder can be excluded and minimized, repectively. In addition, Weischenberg et al. reported that on-site Coulomb interactions are not required to reproduce the experimental ANC in Fe [30]. In fact, the *c/a* ratio dependence of $b_{int}$ is qualitative agreement with the trend of $\sigma_{xy}^{int}$ without considering on-site Coulomb interaction. Therefore, the observed discrepancy between the $\alpha_{xy}^A$ and the $\alpha_{xy}^{int}$ in Fe films is primarily attributed to the extrinsic contribution.

Below, we explain why the intrinsic contribution is dominant in AHC, whereas the extrinsic contribution become more significant in the ANC. The relation between ANC ($\alpha_{xy}$) and AHC ($\sigma_{xy}$) is expressed as Mott's relation,

$$\alpha_{xy} = -\frac{\pi^2}{3}\frac{k_B^2 T}{|e|}\frac{d\sigma_{xy}}{d\mu}, \qquad (6)$$

where $k_B$ is the Boltzmann constant. The $\alpha_{xy}$ relates with the energy derivative of the $\sigma_{xy}$, indicating independence from the magnitude of $\sigma_{xy}$. For example, the theoretical calculation for $L1_0$-$Co_{50}Pt_{50}$ without on-site Coulomb interaction predicts a negligible $\sigma_{xy}^{int}$ ($\approx$ 0 S/cm) at the $E_F$, while the $\alpha_{xy}^{int}$ shows a large value ($\approx$ 4 A/(m·K)) [34]. Pu et al. have analyzed the intrinsic and extrinsic contributions of $\alpha_{xy}$ using Mott' relation [56]. As mentioned in the example, the extrinsic $\frac{d\sigma_{xy}}{d\mu}$ may significantly influence the $\alpha_{xy}$ even if the extrinsic $\sigma_{xy}$ itself is negligibly small. This relation further suggests that the influence of the extrinsic contribution on the $\alpha_{xy}^A$ cannot be simply described.

Finally, in the sample with strain gradients, the $b_{int}$ and the $\alpha_{xy}^A$ show abrupt changes, which are inconsistent with the $\sigma_{xy}^{int}$ and the $\alpha_{xy}^{int}$ due to strain gradients. As shown in the inset of Fig. 1(d), since the Fe film has slight strain gradients at *c/a* = 1.0108, the $b_{int}$ was inconsistent with that of the $\sigma_{xy}^{int}$.

In conclusion, we experimentally and theoretically studied the influence of strain to the AHE and the ANE in Fe epitaxial film grown on MgO and MAO substrates with various sputtering conditions. The intrinsic AHC evaluated by the scaling law analysis clearly changes with the variation of the *c/a* ratio. A qualitative agreement between the experimental and theoretical AHC as a function of *c/a* ratio was obtained, indicating that the effect of strain-induced modulation in the Berry curvature on the AHE in the Fe epitaxial films. In contrast, the ANC $\alpha_{xy}^A$ exhibits negative values irrespective of the *c/a* ratio, in clear disagreement with the first-principles calculations, even in their sign. This discrepancy indicates a non-negligible extrinsic contribution to $\alpha_{xy}^A$ in the present Fe films, in contrast to the predominant intrinsic nature of the AHC. These results highlight a crucial difference in the physical origin between the AHC and the ANC in Fe; the AHC is dominated by intrinsic mechanisms, whereas the ANC is largely influenced by the extrinsic contribution.

**Acknowledgements:**

This work was supported by JST ERATO "Magnetic Thermal Management Materials" [Grant No. JPMJER2201], JST CREST "Exploring Innovative Materials in Unknown Search Space" [Grant No. JPMJCR21O1], A-STEP (Stage II, Full-scale type) (Grant No. JPMJTR253A), Grants-in-Aid for Scientific Research (Grants No. 24K00932 and No. 26K00993), and the NIMS Cooperative Research Program of the National Institute for Materials Science (NIMS). The authors thank S. Kuramochi, N. Kojima, and T. Hiroto for technical supports with the deposition, device fabrication, and XRD measurements, respectively. The first-principles calculation was performed on the Numerical Materials Simulator at NIMS.

**Disclosure statement:**

The authors declare no competing interests.

**DATA AVAILABILITY**

The data supporting the article's findings are openly available [57].

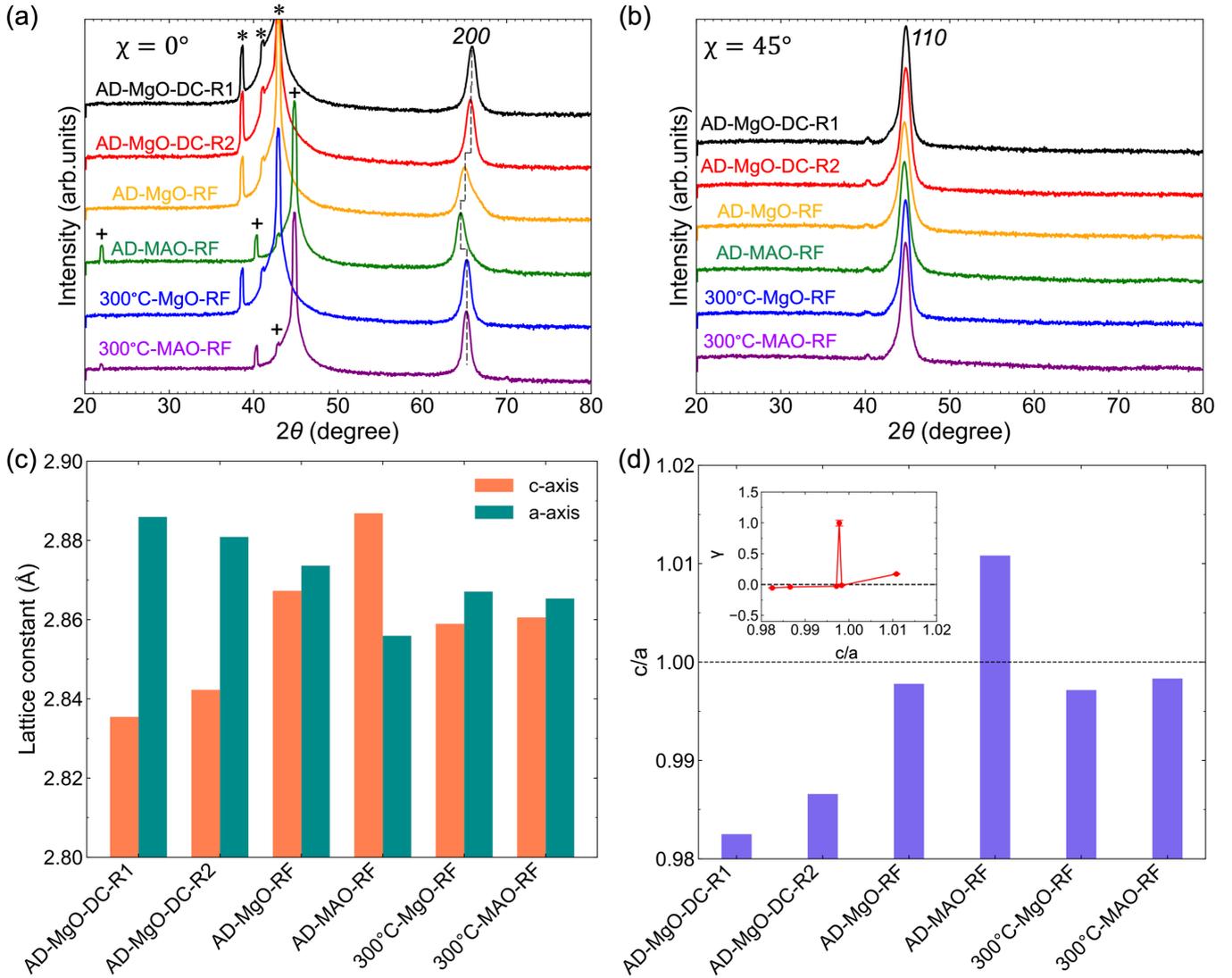

Fig. 1: X-ray diffraction (XRD) results of Fe thin films at (a) $\chi = 0°$ and (b) $\chi = 45°$. (c) The lattice constant derived from 200 and 110 peaks. (d) The *c/a* ratio calculated from the lattice constant along c-axis and a-axis. The inset figure shows the *c/a* ratio dependence of the skewness $\gamma$ obtained 200 peak shapes. The diffraction peaks of the MgO and MAO substrates are denoted by the symbols (* and +) in part (a), respectively.

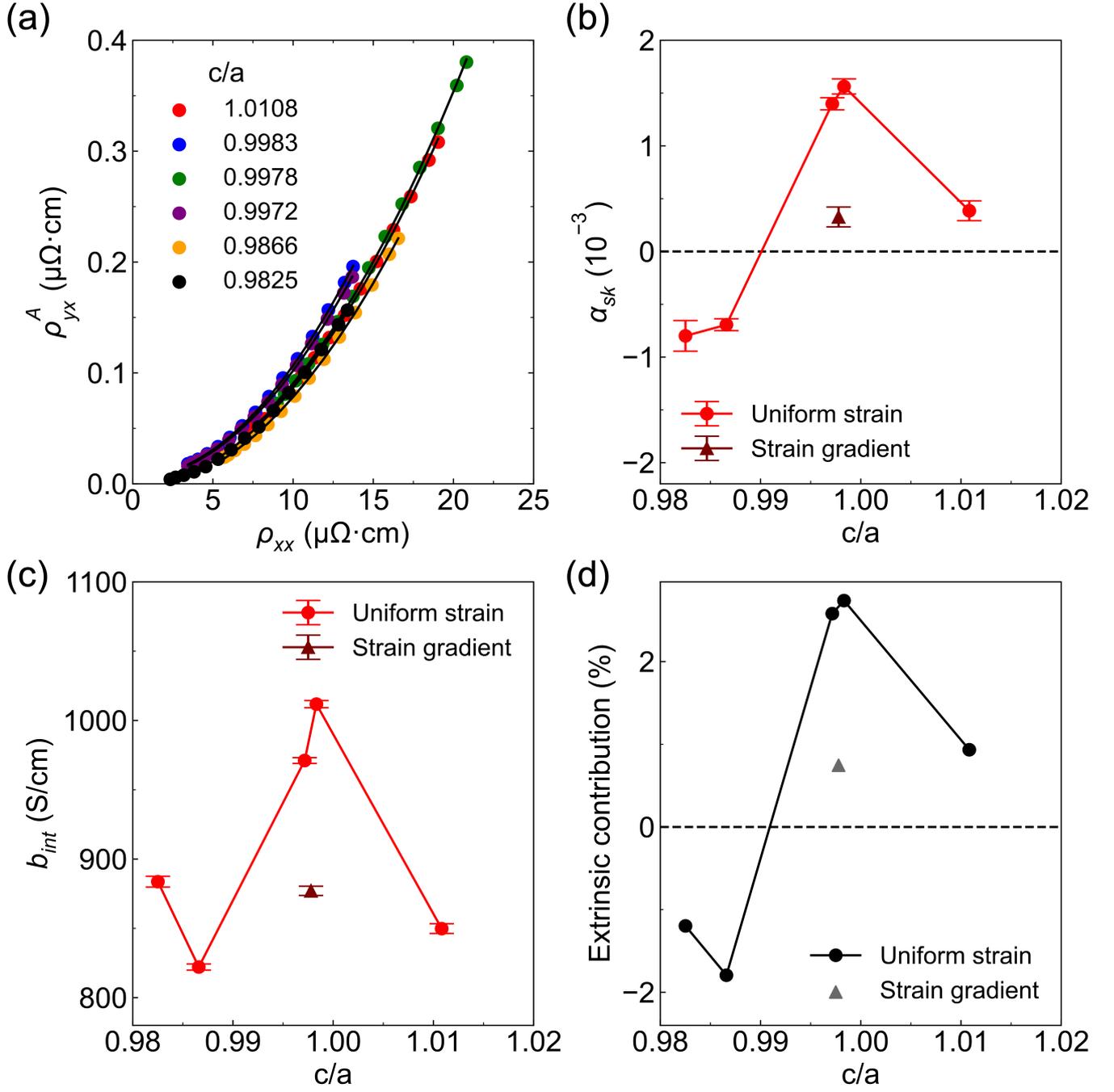

Fig. 2: (a) Fitting results of a scaling analysis ($\rho_{yx}^A = \alpha_{sk}\rho_{xx0} + b_{int}\rho_{xx}^2$) at each c/a ratio. The fitting line is indicated by black solid lines. *c/a* ratio dependence of (b) the extrinsic contribution $\alpha_{sk}$, (c) the intrinsic AHC $b_{int}$, and (d) the extrinsic contribution ratio $\alpha_{sk}\rho_{xx0} / \rho_{yx}^A$ at 300 K obtained from the scaling analysis.

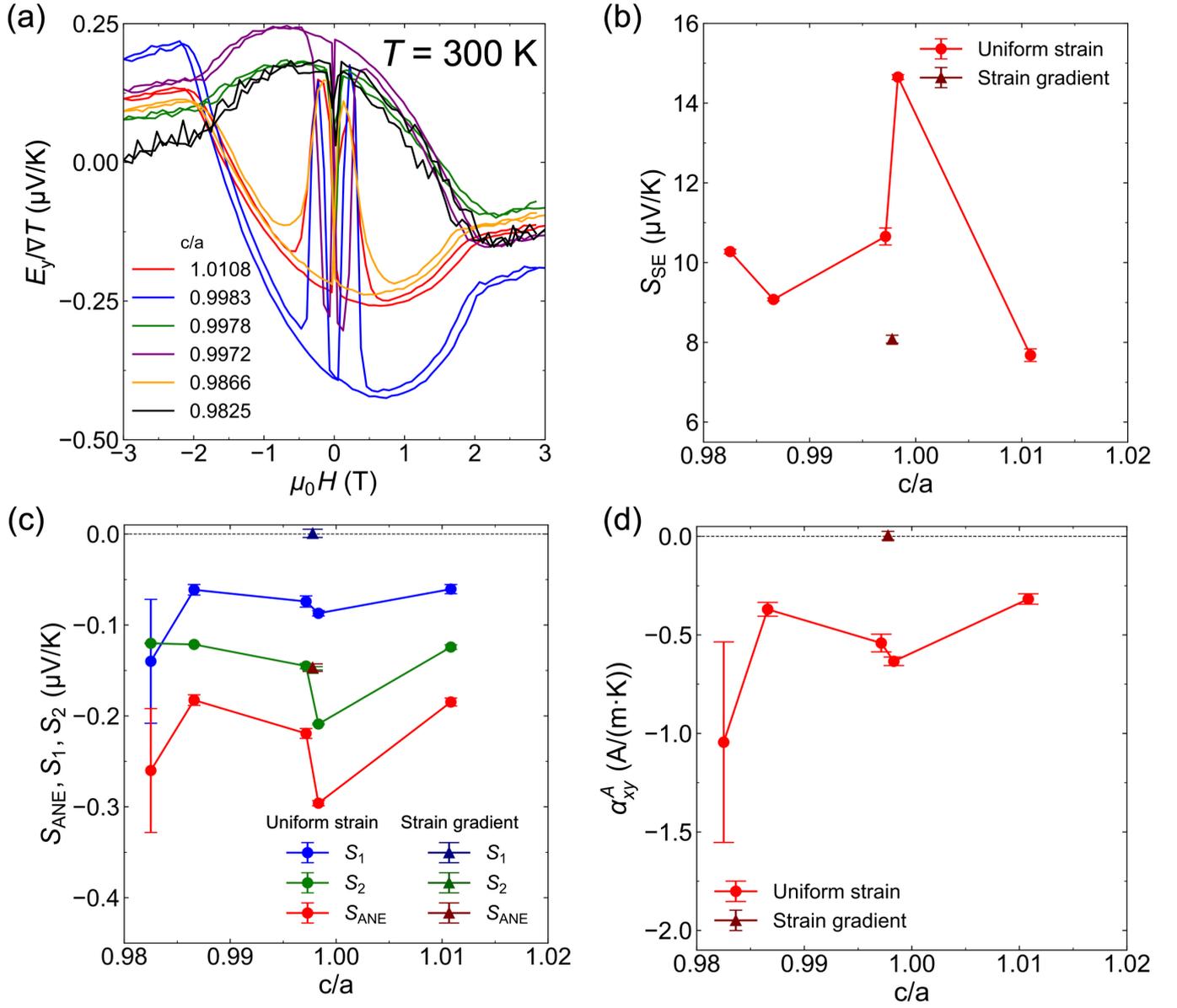

Fig. 3: (a) Magnetic field $H$ dependence of the Nernst electric field normalized by the applied temperature gradient $E_y / \nabla T$. $c/a$ ratio dependence of (b) the Seebeck coefficient $S_{SE}$, (c) the anomalous Nernst coefficient $S_{ANE}$, $S_1$ ($= \rho_{xx}\alpha_{xy}^A$), and $S_2$ ($= -S_{SE}\rho_{yx}^A / \rho_{xx}$), and (d) the ANC $\alpha_{xy}^A$.

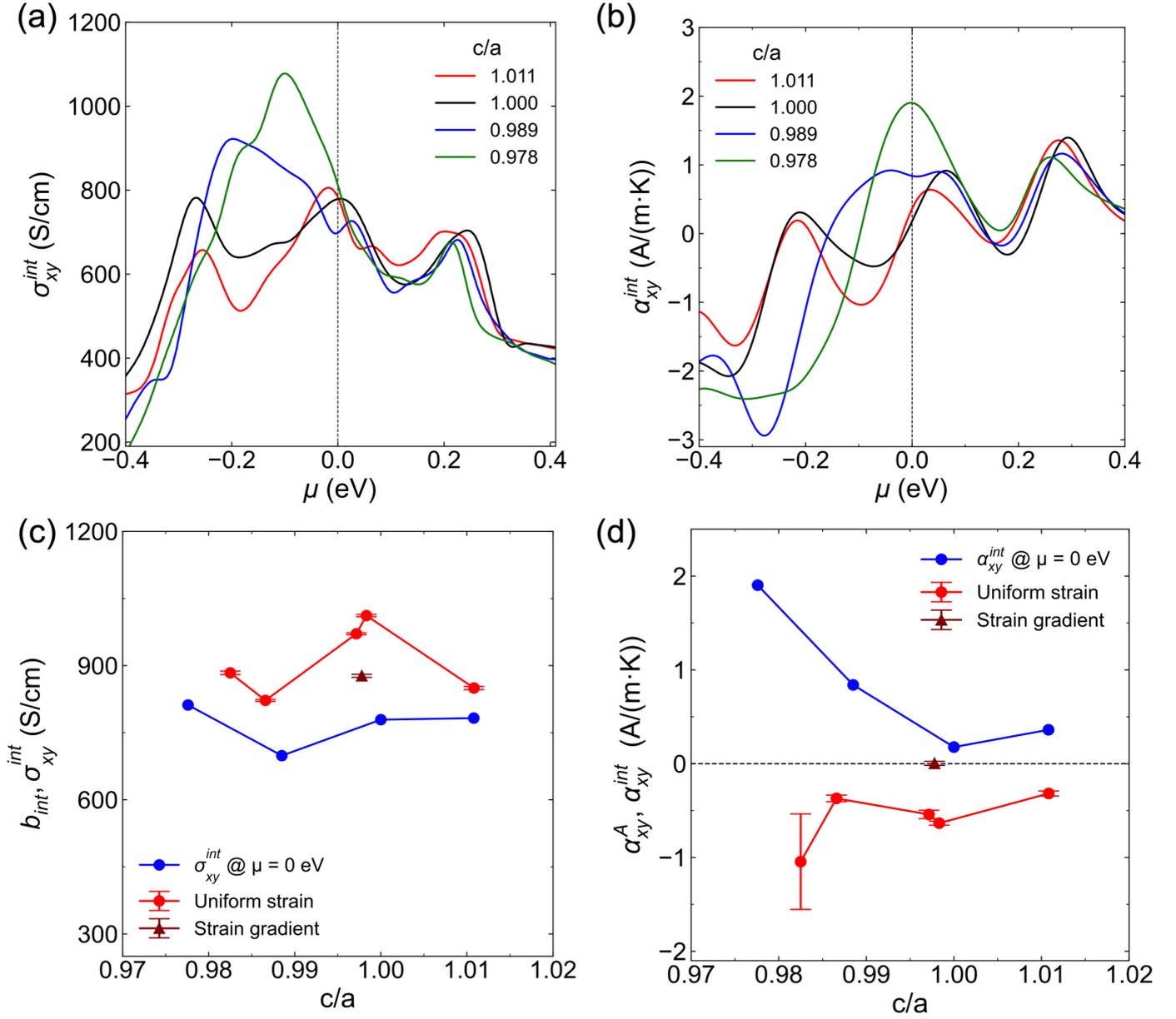

Fig. 4: Theoretical energy dependence of (a) the intrinsic AHC $\sigma_{xy}^{int}$ at 0 K and (b) the intrinsic ANC $\alpha_{xy}^{int}$ at 300 K in Fe. Comparisons of *c/a* ratio dependence for (c) $b_{int}$-$\sigma_{xy}^{int}$ at 0 K and (d) $\alpha_{xy}^{A}$-$\alpha_{xy}^{int}$ at 300 K.